%% file: paper.tex
\documentclass[orivec,a4paper,envcountsect,UKenglish, autoref,10pt]{llncs}

\input{packages}

\input{commands}

\definecolor{orcidlogocol}{HTML}{A6CE39}

\title{In-place implementation of Quantum-Gimli
}

\titlerunning{Quantum-Gimli in-place}

\author{
	Lars Schlieper\thanks{Funded by DFG under Germany's Excellence Strategy - EXC 2092 CASA - 390781972.} \href{https://orcid.org/0000-0002-4870-1012}{\protect\includegraphics[height=\fontcharht\font`B]{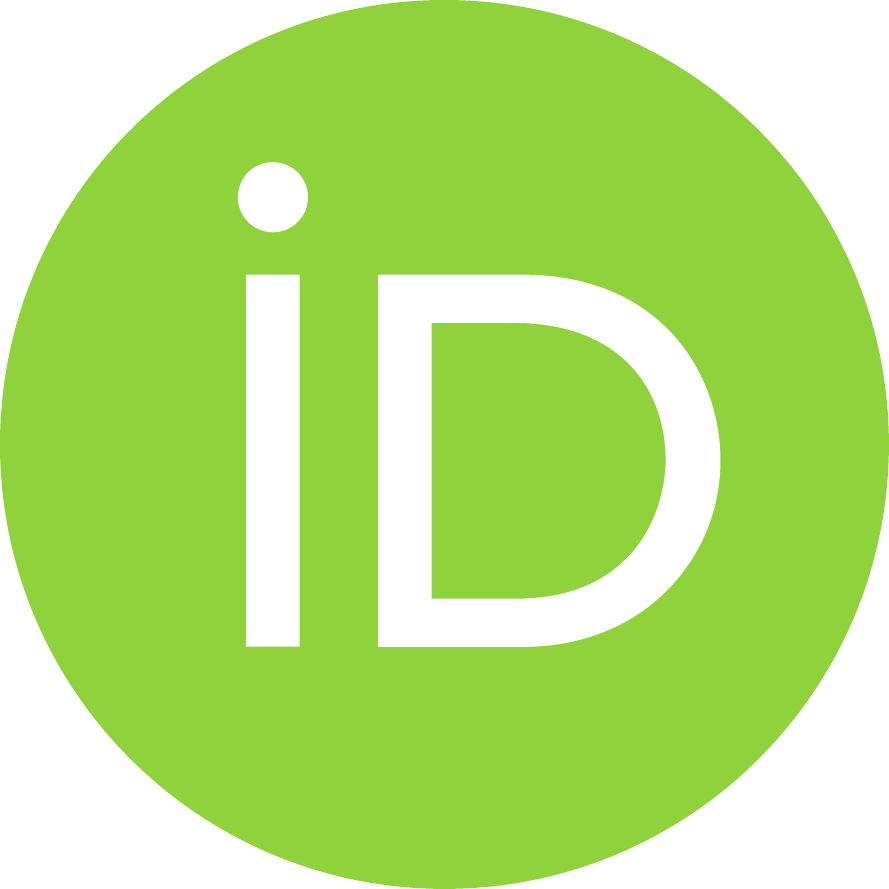}}}
\institute{
	Horst G\"ortz Institute for IT Security \\ 
	Ruhr University Bochum, Germany \\
	\email{\{lars.schlieper\}@rub.de}}

\allowdisplaybreaks
\begin{document}
\pagestyle{plain}

\maketitle
\input{1_introduction.tex}
\input{2_previous.tex}
\input{3_gimli.tex}
\input{4_quantum-gimli.tex}
\input{5_discussion.tex}

\bibliographystyle{plainurl}
\bibliography{literature}
\end{document}

%% file: packages.tex
\usepackage{hyperref}
\usepackage{subcaption}
\usepackage{url}

\usepackage{comment}
\usepackage[utf8]{inputenc}

\usepackage{verbatim}
\usepackage{graphicx}
\usepackage{xcolor}
\usepackage{float}

\usepackage{amsthm}
\usepackage{amsfonts,amsmath,amssymb}
\usepackage[capitalise,noabbrev]{cleveref}
\usepackage{braket}
\usepackage{mathtools}

\usepackage{array}
\usepackage{multirow}
\usepackage[ruled,vlined,linesnumbered]{algorithm2e}
\usepackage{csquotes}
\usepackage{tabularx}
\usepackage{algorithm2e}
\usepackage{algorithmicx}
\usepackage[noend]{algpseudocode}

\usepackage{circuitikz}
\usepackage{tikz}
\usepackage{qcircuit}
\usepackage{mathtools}
\usepackage{pgfplots}

\usepackage{calc}

\usetikzlibrary{
    calc,
    trees,
    positioning,
    arrows,
    chains,
    shapes.geometric, 
    decorations.pathreplacing,
    decorations.pathmorphing,
    shapes, 
    matrix,
    shapes.symbols
    }

%% file: commands.tex
\newlength{\strutdepth}%
\newcommand{\notes}[3]{
	\noindent{%
		\color{#1}{[#3]}\color{#1}}%
	\strut\vadjust{\kern-\strutdepth%
		\vtop to \strutdepth{%
			\baselineskip\strutdepth%
			\vss\llap{{\large\color{#1}\textbf{#2}\quad\color{black}}}\null%
		}%
	}%
}

\newcommand{\NOT}{\textnormal{NOT}}
\newcommand{\CNOT}{\textnormal{CNOT}}
\newcommand{\CCNOT}{\textnormal{CCNOT}}
\newcommand{\SWAP}{\textnormal{SWAP}}

\newtheorem*{lemma*}{Lemma}

%% file: 1_introduction.tex

\begin{abstract}
	We present an in-place implementation of the cryptographic permutation \textsc{Gimli}, a NIST round 2 candidate for lightweight cryptography, and provide an upper bound for the required quantum resource in depth and gate-counts. 
	In particular, we do not use any ancilla qubits and the state that our circuit produces is not entangled with any input.
	This offers further freedom in the usability and allows for a widespread use in different applications in a plug-and-play manner.

	\keywords{Quantum Algorithm, Implementation, Permutation, \textsc{Gimli}, In-Place, Circuit, Reducing/Minimize Qubits.}
\end{abstract}

%
%

\section{Introduction}\label{sec:introduction}

In recent years, the realization of quantum computers has made great progress \cite{Comparison}.
With the launch of the NIST post quantum competition quantum computers have moved even more into the focus of security considerations. One of the subjects being considered in research is the limits of quantum attacks on existing cryptosystems.
Recently there have been many results concerning Grover attacks on symmetric ciphers, especially on lightweight ciphers.
This includes related implementations and resource estimates of the quantum implementation of the ciphers and attacks \cite{SIMON,AES1,SPECK,AES2,AES3}, as well as attempts to adapt classical attacks into the quantum setting \cite{CTQ1,CTQ2,CTQ3,CTQ4}. 
These approaches have in common that they require an (efficient) quantum implementation of the targeted cipher. 

Another reason for such implementations is the connection of quantum computers to a kind of quantum internet \cite{QI2,QI1} and the resulting necessity of encryption procedures. These have to be either efficient implementations of classical encryption schemes or even own quantum encryption schemes which may exploit entanglements.

For many such ciphers permutations are required. These can be implemented in various ways. Two of these possibilities are either as a lookup table or directly as a circuit.
The approach to realize these permutations in a quantum setting by a lookup table is problematic, because for a superposition request to the permutation the table must also allow for an efficient superposition request. 
On the other hand, the circuit approach requires an embedding, which has the advantage that permutations are already reversible.

Apart from the standard reversible embedding of $n$-bit to $n$-bit functions with at least $2n$ bits \cite{Embedding}, Vivek V. et al showed in \cite{Permutation} that any permutation can be implemented with a maximum of $n+1$ bits over the CNT-gate-set (\CNOT, \NOT, \CCNOT), but not necessarily efficiently, namely with polynomial depth and a polynomial number of gates. 

The number of possible permutations alone shows that an efficient implementation over this gate-set is not possible for every permutation.
This raises the question which permutations can be efficiently implemented.

We show in this paper that \textsc{Gimli} \cite{Gimli} is an efficient implementable permutation.
Further more we show that even an in-place implementation is possible.
Therefore the permutation of superpositions is possible without entanglement with "input bits".

%
%

\paragraph{Our contributions:}

To the best of our knowledge we give the first step-by-step reversible in-space implementation of the \textsc{Gimli}-permutation \cite{Gimli} in \textsc{qiskit} \cite{Qiskit}, which is provided by IBM. 
In addition, we give in this case study a polynomial upper-bound for the number of required gates, depth, depending on the number of rounds and the word lengths of \textsc{Gimli}.

Our circuit can be used as a building block or starting point for further applications and research. 
In particular, since there is no entanglement between input and output, this may open up further possibilities.  

Furthermore, the realization of \textsc{Gimli} as a quantum circuit out of reversible gates directly provides a circuit for the inverse of \textsc{Gimli}.

%
%

\paragraph{Organization:}

In \cref{sec:preliminaries} we first introduce the notation and model we use.
In \cref{sec:gimli} we briefly recall the original \textsc{Gimli}-permutation.
In \cref{sec:quantum_gimli} we then build our implementation of the quantum circuit step by step and specify the circuit size.

%
%

%% file: 2_previous.tex

\section{Preliminaries}\label{sec:preliminaries}
Let us first recall some notations. For the classical part we use the same notation as in the original \textsc{Gimli} paper \cite{Gimli}.
We define $\mathcal{W}:=\{0,1\}^{32}$ and use
\begin{itemize}
	\item[-] $a \oplus b $ to denote a bitwise\textbf{ exclusive or} (XOR) of the values $a$ and $b$,
	\item[-] $a \wedge b$ for a bitwise logical \textbf{and} of the values $a$ and $b$,
	\item[-] $a \vee   b$ for a bitwise logical \textbf{or} of the values $a$ and $b$,
	\item[-] $a \lll   k$ for a \textbf{cyclic left shift} of the value $a$ by a shift distance of $k$, and
	\item[-] $a \ll    k$ for a \textbf{non-cyclic shift} (i.e, a shift that is filling up with zero bits) of the value $a$ by a shift distance of $k$.
\end{itemize}
Further we describe our words $w\in\mathcal{W}$ as vectors $w=(w_0,\ldots,w_{31})$. We refer to a $384$-bit state as a $3\times4\times32$-matrix over $\{0,1\}$, or equivalently as a $3\times4$ matrix over words $s_{i,j}\in\mathcal{W}$. 
%
%
The quantum gates used in this paper can be derived from the Clifford+$T$ set and are the \textsc{NOT} gate $X$, Hadamard-gate $H$, Phase shift gates $T,T^\dagger$ to the angles $\frac{\pi}{4},\frac{-\pi}{4}$,
\[
	X:=
	\begin{pmatrix}
		0&1\\
		1&0
	\end{pmatrix}\;,
	\quad
	H:=\frac{1}{\sqrt{2}}
	\begin{pmatrix}
		1&1\\
		1&-1
	\end{pmatrix}\;,
	\quad
	T:=
	\begin{pmatrix}
		1&0\\
		0&\frac{1+i}{\sqrt{2}}
	\end{pmatrix}\;,
	\quad 
	T^\dagger:=
	\begin{pmatrix}
		1&0\\
		0&\frac{1-i}{\sqrt{2}}
	\end{pmatrix}
\]
as well as the multi-qubit gates \textsc{CNOT} (controlled \textsc{NOT}) and \textsc{CCNOT} (Toffoli),
\[
	\textsc{CNOT}:=
	\begin{pmatrix}
		1&0&0&0\\
		0&1&0&0\\
		0&0&0&1\\
		0&0&1&0
	\end{pmatrix}
	\sim	
	\raisebox{0.45cm}{\Qcircuit @C=1.0em @R=1.0em @!R { 
			& \ctrl{1}	&\qw\\
			& \targ		&\qw
	}}
	\;,\quad
	\textsc{CCNOT}:=
	\begin{pmatrix}
		1&0&0&0&0&0&0&0\\
		0&1&0&0&0&0&0&0\\
		0&0&1&0&0&0&0&0\\
		0&0&0&1&0&0&0&0\\
		0&0&0&0&1&0&0&0\\
		0&0&0&0&0&1&0&0\\
		0&0&0&0&0&0&0&1\\
		0&0&0&0&0&0&1&0
	\end{pmatrix}
	\sim	
	\raisebox{0.75cm}{\Qcircuit @C=1.0em @R=1.0em @!R { 
			& \ctrl{2}	&\qw\\
			& \ctrl{1}	&\qw\\
			& \targ		&\qw
	}}\;,
\]
\noindent which can be constructed in-place with the gate above, as seen in \cref{fig:circ:toffoli}. 

\begin{figure}[h]
	\centering
	\scalebox{1}{
		\mbox{
			\Qcircuit @C=1.0em @R=1.0em @!R { 
				& \ctrl{2}	&\qw	&   && \qw 		  & \qw 	& \qw 				& \ctrl{2}  & \qw		& \qw		& \qw		         & \ctrl{2} & \qw      & \ctrl{1} & \gate{T}		 & \ctrl{1} 	& \qw\\	
				& \ctrl{1}	&\qw	& = && \qw 		  & \ctrl{1}& \qw 				& \qw 		& \qw		& \ctrl{1}	& \qw                & \qw   	& \gate{T} & \targ    & \gate{T^\dagger} & \targ 		& \qw\\
				& \targ		&\qw	&   && \gate{H}   & \targ   & \gate{T^\dagger}  & \targ 	& \gate{T}  & \targ 	& \gate{T^\dagger}   & \targ 	& \gate{T} & \gate{H} & \qw			     & \qw   		& \qw
			}
	}}
	\caption{In-place implementation of Toffoli via $H$, $T$, $T^\dagger$ and \CNOT-gates.}\label{fig:circ:toffoli}
\end{figure}
%
\noindent Further we use the \SWAP-operation,
\[
	\textsc{SWAP}:=
	\begin{pmatrix}
		1&0&0&0\\
		0&0&1&0\\
		0&1&0&0\\
		0&0&0&1
	\end{pmatrix}
	\sim	
	\raisebox{0.45cm}{\Qcircuit @C=1em @R=1em {
			& \qswap      & \qw & \raisebox{-2.3em}{$\sim$} &  & \ctrl{1}	& \qw & \targ     & \qw & \ctrl{1}  & \qw & \raisebox{-2.3em}{$\sim$}	&  & \qw & \link{1}{-1}  & \qw &\\
			& \qswap \qwx & \qw &  				 			&  & \targ   	& \qw & \ctrl{-1} & \qw & \targ     & \qw &                       		&  & \qw & \link{-1}{-1} & \qw & 
		}
	}\;.
\]

We consider in this paper quantum-circuits over the set of gates above, under the standard assumptions of full-parallelism, full-connectivity and we do not consider errors. 
Associated with this we do not count \SWAP-operations and take them as free, instead we will re-label the qubits accordingly as in \cite{AES2}. 

%
%

%% file: 3_gimli.tex

\newpage
\section{\textsc{Gimli}}\label{sec:gimli}

\textsc{Gimli} \cite{Gimli} is a $384$-bit permutation, candidate of the second round of NISTs lightweight cryptography competition and designed to achieve high security with high performance across a broad range of platforms.
This cryptographic primitive is suitable for many different applications, e.g. collision-resistant hashing, preimage-resistant hashing, message authentication, and message encryption.

Let us briefly recall \textsc{Gimli} (Algorithm \ref{alg:gimli}).
It is a round-based permutation on a $384$-bit state $s=(s_{i,j})\in\mathcal{W}^{3\times4}$.
Each of the $24$ rounds is a sequence of at most three operations: 
\begin{itemize}
	\item[-] a non-linear layer, where a $96$-bit SP-Box is applied to each column;
	\item[-] in every second round, a linear mixing layer;
	\item[-] in every fourth round, an addition of a constant and the round-number. 
\end{itemize}

\begin{algorithm}[H]
	\DontPrintSemicolon
	\SetAlgoLined
	\SetKwInOut{Input}{Input}\SetKwInOut{Output}{Output}
	
	\SetKwComment{COMMENT}{$\triangleright$\ }{}%
	
	\Input{ $s=(s_{i,j})\in\mathcal{W}^{3\times4}$}
	\Output{\textsc{Gimli}$(s)\in\mathcal{W}^{3\times4}$ }
	
	\Begin{
		\For{r from $24$ downto $1$ inclusive}{
			\For(\COMMENT*[f]{SP-Box}){j from $0$ to $3$ inclusive}{ 
				$x\leftarrow s_{0,j}\lll 24$\;
				$y\leftarrow s_{1,j}\lll 9$ \;
				$z\leftarrow s_{2,j}$ 		\;
				$s_{2,j}\leftarrow x	\oplus(z\ll1)	\oplus	((y\wedge z)\ll2)$ \;
				$s_{1,j}\leftarrow x	\oplus y		\oplus	((x\vee z)\ll1)$ \;
				$s_{0,j}\leftarrow y	\oplus z		\oplus	((x\wedge y)\ll3)$ \;
			}
			\If(\COMMENT*[f]{Small-Swap}){$r\equiv 0 \mod 4$}{
				$s_{0,0}s_{0,1}s_{0,2}s_{0,3}\leftarrow s_{0,1}s_{0,0}s_{0,3}s_{0,2}$ \;
			}
			\If(\COMMENT*[f]{Big-Swap}){$r\equiv 2 \mod 4$}{
				$s_{0,0}s_{0,1}s_{0,2}s_{0,3}\leftarrow s_{0,2}s_{0,3}s_{0,0}s_{0,1}$ \;
			}
			\If(\COMMENT*[f]{Add constant}){$r\equiv 0 \mod 4$}{
				$s_{0,0}\leftarrow s_{0,0}\oplus c \oplus r$ \;
			}	
		}
		\KwRet{$s$}
		
	}
	\caption{\textsc{Gimli}}\label{alg:gimli}
\end{algorithm}

\begin{remark}
	\textsc{Gimli} appears to be easily scalable in the number of rounds $r$ and the length $\ell$ of the words $s$ in $\mathcal{W}$, where scaling may affect the security.
\end{remark}

%
%

%% file: 4_quantum-gimli.tex

\section{Quantum \textsc{Gimli}}\label{sec:quantum_gimli}
In this section we give the description of our in-place quantum circuit for the \textsc{Gimli} permutation. 
Let us begin by subdividing \textsc{Gimli} into even smaller pieces.

The linear parts of \textsc{Gimli} are two different \SWAP\ operations, called \textsc{Small-SWAP} and \textsc{Big-SWAP}, and a \textsc{XOR} with a constant and the round-number.
Samuel Jaques et al. describe in \cite{PLU} how every invertible linear function can be implemented efficiently in-place by using the numerical procedure of PLU 
decomposition.
In case of the linear functions of \textsc{Gimli} it is even easier to implement them.
The \textsc{XOR} with a constant and the round-number can be hard-wired at the necessary points via \textsc{NOT}-gates, since they are known in advance.
The two different \SWAP\ operations can either be implemented by \textsc{SWAP}-gates or by relabelling.
While using the relabelling technique we have to pay attention to the new labels of the qubits and so we have to be careful to use the correct bits in the further calculation.
The underlying bit-permutation can be easily calculated and can be created with the algorithm in the supplementary material.
At the end either the new labels have to be taken into account for further computations or we have to add a \textsc{SWAP}-layer to swap the qubits via \textsc{SWAP}-gates back in the correct order.

The non-linear part of \textsc{Gimli}, the \textsc{SP-Box}, is a bit more tricky.
The \textsc{SP-Box} works on three inputs words $x,y,z\in\mathcal{W}$ ($96$-bits) and can again be split into three parts:
\begin{enumerate}
	\item\label{Part1} a cyclic shift of $x$ and $y$ by $24$ and $9$
	\begin{align*}
	x\leftarrow& x\lll 24 \\
	y\leftarrow& y\lll  9 \;.
	\end{align*}
	
	\item\label{Part2} three parallel updates of $x,y$ and $z$ via a T-function with non-cyclic shifts as part of the calculation
	\begin{align}
	x\leftarrow& x	\oplus \makebox[1.3cm][c]{($z\ll1$)}	\oplus	((y \wedge z) \ll 2) \label{x}\\
	y\leftarrow& x	\oplus \makebox[1.3cm][c]{$y$}			\oplus	((x \vee   z) \ll 1) \label{y}\\
	z\leftarrow& y	\oplus \makebox[1.3cm][c]{$z$}			\oplus	((x \wedge y) \ll 3) \label{z}\;.
	\end{align}
	
	\item\label{Part3} a Swap of $x$ and $z$ 
	\begin{align*}
	x\leftarrow& z \\
	z\leftarrow& x \;.
	\end{align*}
\end{enumerate}

The parts \ref{Part1} and \ref{Part3} can also be done by relabelling similar to the \SWAP s. It remains to show how part \ref{Part2} (the T-function) can be efficiently implemented in-place.
First we notice that to achieve an in-place implementation we can \textbf{not} simply use the "classic" universal embedding of the T-function, update $x,y$ and $z$ as in \textsc{Gimli} via an additional register, nor calculate $x,y$ and $z$ one after the other, since they are dependent on each others non-updated value and we do \textbf{not} want to use extra space for this calculations.

\begin{figure}[H]
	\centering
	\scalebox{1}{
	\mbox{
		\Qcircuit @C=1em @R=0.5em @! {
			\lstick{\ket{x_k}}		&	\qw			&	\qw\barrier[-0.4cm]{11}	&	\ctrl{1}&	\qw					&	\qw			&	\qw\barrier[-0.3cm]{11}	&	\targ		&	\targ		&\rstick{\ket{x_k'}}\qw\\
			\lstick{\ket{y_k}}		&	\ctrl{1}	&	\qw						&	\targ	&	\qw					&	\targ		&	\gate{X}				&	\qw			&	\qw			&\rstick{\ket{y_k'}}\qw\\
			\lstick{\ket{z_k}}		&	\targ		&	\targ					&	\qw		&	\qw					&	\qw			&	\qw						&	\qw			&	\qw			&\rstick{\ket{z_k'}}\qw\\
			\lstick{\ket{x_{k+1}}}	&	\qw			&	\qw						&	\qw		&	\gate{X}			&	\ctrl{-2}	&	\gate{X}				&	\qw			&	\qw			&\rstick{\ket{x_{k+1}}}\qw\\
			\lstick{\ket{y_{k+1}}}	&	\qw			&	\qw						&	\qw		&	\qw					&	\qw			&	\qw						&	\qw			&	\qw			&\rstick{\ket{y_{k+1}}}\qw\\
			\lstick{\ket{z_{k+1}}}	&	\qw			&	\qw						&	\qw		&	\gate{X}			&	\ctrl{-4}	&	\gate{X}				&	\qw			&	\ctrl{-5}	&\rstick{\ket{z_{k+1}}}\qw\\
			\lstick{\ket{x_{k+2}}}	&	\qw			&	\qw						&	\qw		&	\qw					&	\qw			&	\qw						&	\qw			&	\qw			&\rstick{\ket{x_{k+2}}}\qw\\
			\lstick{\ket{y_{k+2}}}	&	\qw			&	\qw						&	\qw		&	\qw					&	\qw			&	\qw						&	\ctrl{-7}	&	\qw			&\rstick{\ket{y_{k+2}}}\qw\\
			\lstick{\ket{z_{k+2}}}	&	\qw			&	\qw						&	\qw		&	\qw					&	\qw			&	\qw						&	\ctrl{-8}	&	\qw			&\rstick{\ket{z_{k+2}}}\qw\\
			\lstick{\ket{x_{k+3}}}	&	\qw			&	\ctrl{-7}				&	\qw		&	\qw					&	\qw			&	\qw						&	\qw			&	\qw			&\rstick{\ket{x_{k+3}}}\qw\\
			\lstick{\ket{y_{k+3}}}	&	\qw			&	\ctrl{-8}				&	\qw		&	\qw					&	\qw			&	\qw						&	\qw			&	\qw			&\rstick{\ket{y_{k+3}}}\qw\\
			\lstick{\ket{z_{k+3}}}	&	\qw			&	\qw						&	\qw		&	\qw					&	\qw			&	\qw						&	\qw			&	\qw			&\rstick{\ket{z_{k+3}}}\qw\\
			&	z'_k		&							&			&	\hspace*{1cm}y'_k	&				&							&				&	x'_k		&			&
	}	}}
	\caption{Circuit for updating $x_k,y_k$ and $z_k$ "parallel" in depth $5$ (T$_k$-function). With $(x_k,y_k,z_k)=(0,0,0)$ for $k\ge32$.}\label{fig:circ:gimli_T_k}
\end{figure}
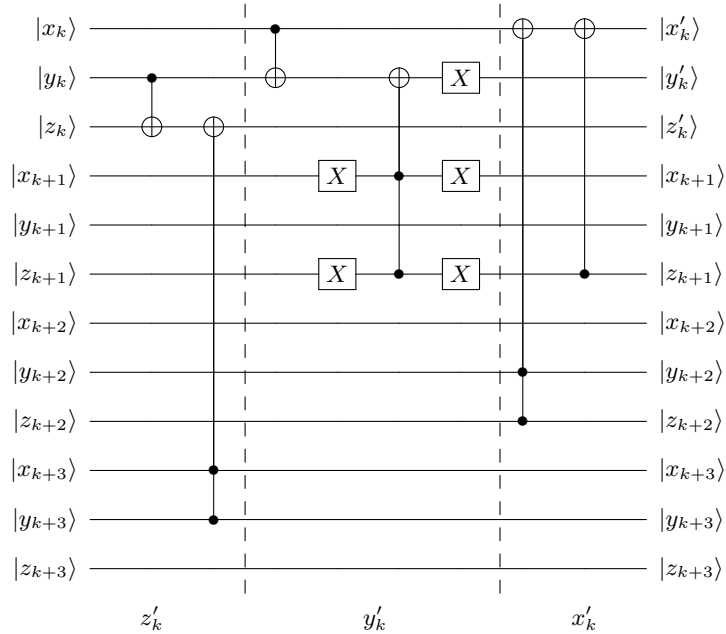

We instead use a bitwise approach and exploit the fact, that for each bit $x_k,y_k$ and $z_k$ of $x,y$ and $z$ no bits with a lower index are used to calculate the update and that by a non-cyclic shift the vector is refilled with zeros. We hence compute $x,y$ and $z$ bit by bit in such an order that the updated bits are no longer necessary for further updates of other bits.
To make our approach more precise, let us first take a look at the update of the individual bits of $x,y$ and $z$ in \cref{x,,y,z}.
The individual bits are computed as 
\begin{align}
x_k\leftarrow& x_k	\oplus \makebox[1cm][c]{$z_{k+1}$}	\oplus ( y_{k+2} \cdot  z_{k+2}) \label {x_k} \\
y_k\leftarrow& x_k	\oplus \makebox[1cm][c]{$y_k$}		\oplus ((x_{k+1} \oplus       1) \cdot  (z_{k+1} \oplus 1 ) ) \oplus 1 \label{y_k} \\
z_k\leftarrow& y_k	\oplus \makebox[1cm][c]{$z_k$}		\oplus ( x_{k+3} \cdot  y_{k+3}) \label {z_k} \;,
\end{align}
where we used De Morgan's laws with \mbox{$A\vee B= \overline{\overline{A}\wedge\overline{B}}$} for the computation of $y_k$ and define $(x_j,y_j,z_j)=(0,0,0)$ for $j>31$.
If we can compute $x_k,y_k$ and $z_k$ "parallel" without use of any ancilla bits now, we can compute $x,y$ and $z$ inductively from $k=0$ to $k=31$ in-place. 
Indeed, this can be done by the circuit seen in \cref{fig:circ:gimli_T_k}, where $x'_k,y'_k$ and $z'_k$ represent the updated $x_k,y_k$ and $z_k$ as shown in \cref{x_k,,y_k,z_k}.

\begin{figure}[h]
	\centering
	\scalebox{1}{
	\mbox{
		\Qcircuit @C=1.0em @R=0.5em @!R {
			\lstick{\ket{(x,y,z)_0}}	 & \multigate{3}{\text{T}_0}& \qw			 			& \qw 	& \ldots && \qw						& \qw &	  &&&			\\
			\lstick{\ket{(x,y,z)_1}}	 & \ghost{\text{T}_0}		& \multigate{3}{\text{T}_1} & \qw 	& \ldots && \qw 					& \qw &	  &&&			\\
			\lstick{\ket{(x,y,z)_2}}	 & \ghost{\text{T}_0}		& \ghost{\text{T}_1}		& \qw 	& \ldots && \qw 					& \qw &	  &&& \lstick{\ket{x}} &	\multigate{2}{\text{T}}  & \qw 		\\
			\lstick{\ket{(x,y,z)_3}}	 & \ghost{\text{T}_0}		& \ghost{\text{T}_1}		& \qw 	& \ldots && \qw 					& \qw &	= &&& \lstick{\ket{y}} &	\ghost{\text{T}}		 & \qw		\\
			\lstick{\ket{(x,y,z)_4}}	 & \qw 						& \ghost{\text{T}_1}		& \qw 	& \ldots && \qw 					& \qw &	  &&& \lstick{\ket{z}} &	\ghost{\text{T}}		 & \qw		\\
			\lstick{\vdots}				 &							&							& \vdots& \ddots && \vdots					&	  &	  &&&			\\
			\lstick{\ket{(x,y,z)_{31}}}	 & \qw 						& \qw 						& \qw 	& \ldots && \gate{\text{T}_{31}} 	& \qw &	  &&&			
		}
	}}
	\caption{T Circuit.}\label{fig:circ:gimli_T}
	\vspace*{-1em}
\end{figure}
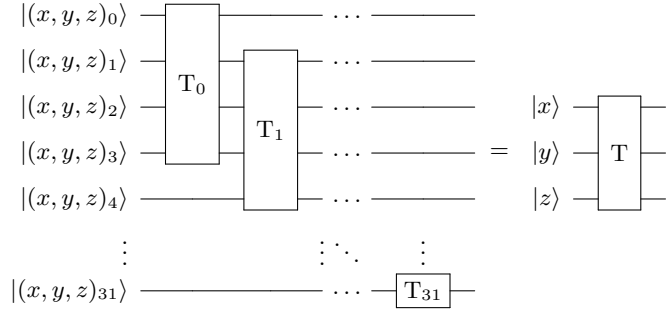

\newpage
\noindent For $k\ge29$ the \textsc{CNOT}s and \textsc{CCNOT}s with missing control-bits are adjusted. For $k=31$ the \textsc{NOT} on $y_k$ is omitted, since it would be cancelled with the \textsc{CCNOT} only controlled with ones.
From here on we can build up the whole circuit of our \textsc{Gimli} in-place implementation. First we get the full T-function as seen in \cref{fig:circ:gimli_T}.

\begin{figure}[h]
	\centering
	\scalebox{1}{
	\mbox{
		\Qcircuit @C=1.0em @R=0.5em @!R { 
			&								& 					& \text{Add constant}&\\
			\lstick{\ket{s_{0,0}}}  & \multigate{2}{\text{SP-Box}} 	& \qswap			& \gate{\text{XOR c}}& \multigate{2}{\text{SP-Box}}		& \multigate{2}{\text{SP-Box}}	& \qswap	& \qw		& \multigate{2}{\text{SP-Box}}	& \qw	\\
			\lstick{\ket{s_{1,0}}}  & \ghost{\text{SP-Box}}			& \qw\qwx			& \qw				 & \ghost{\text{SP-Box}}			& \ghost{\text{SP-Box}}			& \qw\qwx	& \qw		& \ghost{\text{SP-Box}}			& \qw	\\	
			\lstick{\ket{s_{2,0}}}  & \ghost{\text{SP-Box}}			& \qw\qwx			& \qw				 & \ghost{\text{SP-Box}}			& \ghost{\text{SP-Box}}			& \qw\qwx	& \qw		& \ghost{\text{SP-Box}}			& \qw	\\	
			\lstick{\ket{s_{0,1}}}  & \multigate{2}{\text{SP-Box}}	& \qswap\qwx		& \qw				 & \multigate{2}{\text{SP-Box}}		& \multigate{2}{\text{SP-Box}}	& \qw\qwx	& \qswap 	& \multigate{2}{\text{SP-Box}}	& \qw	\\	
			\lstick{\ket{s_{1,1}}}  & \ghost{\text{SP-Box}}			& \qw				& \qw				 & \ghost{\text{SP-Box}}			& \ghost{\text{SP-Box}}			& \qw\qwx	& \qw\qwx	& \ghost{\text{SP-Box}}			& \qw	\\	
			\lstick{\ket{s_{2,1}}}  & \ghost{\text{SP-Box}}			& \qw				& \qw				 & \ghost{\text{SP-Box}}			& \ghost{\text{SP-Box}}			& \qw\qwx	& \qw\qwx	& \ghost{\text{SP-Box}}			& \qw	\\	
			\lstick{\ket{s_{0,2}}}  & \multigate{2}{\text{SP-Box}}	& \qswap			& \qw				 & \multigate{2}{\text{SP-Box}}		& \multigate{2}{\text{SP-Box}}	& \qswap\qwx& \qw\qwx	& \multigate{2}{\text{SP-Box}}	& \qw	\\	
			\lstick{\ket{s_{1,2}}}  & \ghost{\text{SP-Box}}			& \qw\qwx			& \qw				 & \ghost{\text{SP-Box}}			& \ghost{\text{SP-Box}}			& \qw		& \qw\qwx	& \ghost{\text{SP-Box}}			& \qw	\\
			\lstick{\ket{s_{2,2}}}  & \ghost{\text{SP-Box}}			& \qw\qwx			& \qw				 & \ghost{\text{SP-Box}}			& \ghost{\text{SP-Box}}			& \qw		& \qw\qwx	& \ghost{\text{SP-Box}}			& \qw	\\	
			\lstick{\ket{s_{0,3}}}  & \multigate{2}{\text{SP-Box}}	& \qswap\qwx		& \qw				 & \multigate{2}{\text{SP-Box}}		& \multigate{2}{\text{SP-Box}}	& \qw		& \qswap\qwx& \multigate{2}{\text{SP-Box}}	& \qw	\\	
			\lstick{\ket{s_{1,3}}}  & \ghost{\text{SP-Box}}			& \qw				& \qw				 & \ghost{\text{SP-Box}}			& \ghost{\text{SP-Box}}			& \qw		& \qw		& \ghost{\text{SP-Box}}			& \qw	\\	
			\lstick{\ket{s_{2,3}}}  & \ghost{\text{SP-Box}}			& \qw				& \qw				 & \ghost{\text{SP-Box}}			& \ghost{\text{SP-Box}}			& \qw		& \qw		& \ghost{\text{SP-Box}}			& \qw	\\
			&								& \text{Small-Swap}	& 					 &					 				&								& \text{\hspace*{0.5cm}Big-Swap}&		&								&	
		}
	}}
	\caption{4 of 24 Rounds of the \textsc{Gimli}-circuit.}\label{fig:circ:gimli_rounds}
\end{figure}
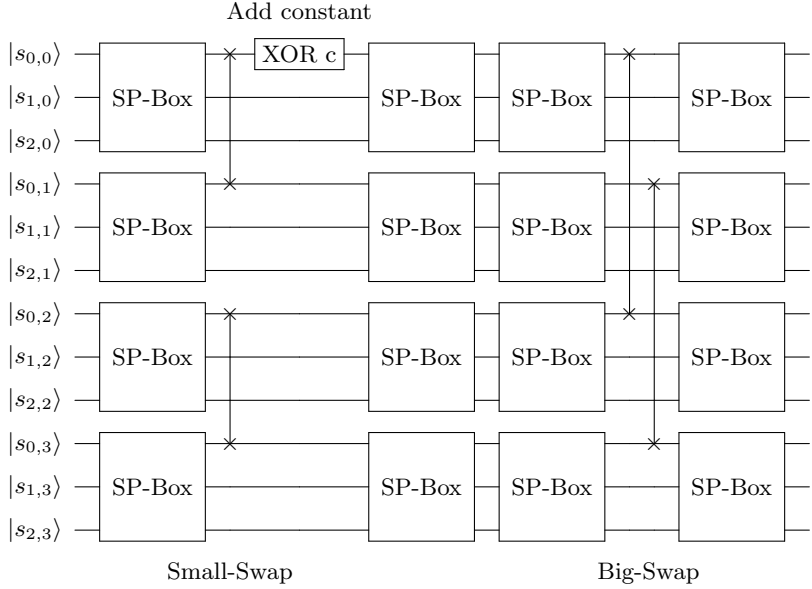

\noindent With this we can build the \textsc{SP-Box} as seen in \cref{fig:circ:gimli_SP-Box}.
\begin{figure}[h]
	\centering
	\scalebox{1}{
	\mbox{
		\Qcircuit @C=1.0em @R=0.5em @!R {
			\lstick{\ket{s_{0,j}}}  & \gate{\lll 24}  & \multigate{2}{\text{T}}		& \qswap 	  & \qw	&	  &&&& \lstick{\ket{s_{0,j}}}  & \multigate{2}{\text{SP-Box}} & \qw	\\
			\lstick{\ket{s_{1,j}}}  & \gate{\lll 9}   & \ghost{\text{T}}			& \qw \qwx	  & \qw	&	= &&&& \lstick{\ket{s_{1,j}}}  & \ghost{\text{SP-Box}}		  & \qw	\\
			\lstick{\ket{s_{2,j}}}  & \qw   		  & \ghost{\text{T}}			& \qswap \qwx & \qw	&	  &&&& \lstick{\ket{s_{2,j}}}  & \ghost{\text{SP-Box}}		  & \qw
		}
	}}
	\caption{Non-linear layer (\textsc{SP-Box}).}\label{fig:circ:gimli_SP-Box}
	\vspace*{-1em}
\end{figure}
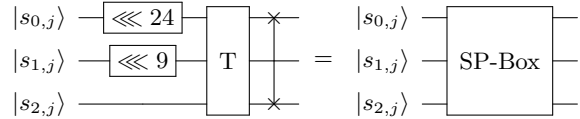\\
The \textsc{SP-Box} together with the discussion above, regarding the linear layers, directly leads to the whole circuit where the circuit-part seen in \cref{fig:circ:gimli_rounds} is repeated $6$ times for $24$ rounds.
\noindent The whole circuit can also be build using algorithm~\ref{alg:circ:q_gimli}.
A python implementation with \textsc{qiskit} of algorithm~\ref{alg:circ:q_gimli} can be found in the supplementary material.

\begin{remark}
	Related to the variables of rounds $r$ and word length $\ell$ the depth of the circuit is upper-bounded by $5\cdot \ell \cdot r + \frac{r}{4}$.
	Here the $5$ comes from the depth of $T_k$ (\cref{fig:circ:gimli_T_k}) and the $\frac{r}{4}$ from the \textsc{XOR} of the key.
\end{remark}

The depth of this circuit for the given parameter $l=32$ and $r=24$ is upper bounded by $3846$, whereas (by shifting parts of the circuit into each other) the circuit produced by algorithm~\ref{alg:circ:q_gimli} has only a depth of $3104$.
This can be further reduced by employing more qubits instead of operating only on the 384 input qubits.

The limitation on $384$ qubits also means that we get the output of the permutation without use of ancilla qubits and entanglement with the input, as would be the case with the "classic" universal embedding, which can provide additional freedom for further use. 

%
%

%% file: 5_discussion.tex

\subsection{Properties}\label{sec:properties}

In the following we list the depth and gate numbers of the circuit generated by algorithm~\ref{alg:circ:q_gimli}.
Here we give the values for a circuit generated with \CCNOT s, as well as the values for the circuit where the \CCNOT s are replaced with the construction from \cref{fig:circ:toffoli}.
Since the $T,T^\dagger$ gates are considered to be the gates with the highest error, we additionally specify the depth of the $T,T^\dagger$ gates for the circuit with replaced \CCNOT s. The results can be seen in \cref{tab:depth_gatecount}, and are optimal in the sense that the internal optimizer of \textsc{qiskit} is not able to improve them.

\begin{table}[h!]
	
	\centering
	\scalebox{0.85}{
	\begin{tabular}{|c|c|c|c|c|c|c|c|c|}\hline
		Circuit				&	Depth	&	Gatecount	&	$X$		&	\CCNOT	&	\CNOT	&	$H$		&	$T$		&	$T$-depth	\\\hline
		With \CCNOT	s		&	3104	&	32739		&	14979	&	8640	&	9120	&	0		&	0		&	0 			\\\hline
		Without \CCNOT s	&	14908	&	153699		&	14979	&	0		&	60960	&	17280	&	60480	&	168			\\\hline
	\end{tabular}}
	\vspace*{0.1cm}
	\caption{Depths and gate numbers of the circuit generated by algorithm~\ref{alg:circ:q_gimli}
		. $T$ and $T^\dagger$ gates are summarized in this table.}\label{tab:depth_gatecount}
\end{table}
\vspace*{-2em}

\begin{remark}
	For varying number of rounds $r$ and word length $\ell$ the depths and gate numbers increase linearly in $r,\ell$.
\end{remark}

For further verification we have programmed a classic 1:1 version of the generated circuitry and let it compete with random input against the original python implementation, 
which produced the same results as \textsc{Gimli} in our 1.000.000 test.
The original implementation of \textsc{Gimli} and the classic version of the circuit can also be found in the supplementary material. 

\newpage
\begin{algorithm}[h]
	\DontPrintSemicolon
	\SetAlgoLined
	\SetKwInOut{Input}{Input}\SetKwInOut{Output}{Output}
	\SetKwComment{COMMENT}{$\triangleright$\ }{}
	
	\Output{Quantum circuit $Q_\textsc{Gimli}$, qubit label dictionary $L$ }
	
	\Begin{
		Define empty quantum circuit $Q_\textsc{Gimli}$ with $384$ qubits.\;
		Define $L=(s_{i,j})_{0\le i< 3, 0\le j< 4}$ as label dictionary.\;
		\For{$r$ from $24$ downto $1$ inclusive}{
			\For(\COMMENT*[f]{SP-Box}){$j$ from $0$ to $3$ inclusive}{
				Relabel qubit registers/update $L$ correspondingly \linebreak $s_{0,j}\lll 24$ and $s_{1,j}\lll 9$.\;
				\For(\COMMENT*[f]{T$_k$}){$k$ from $0$ to $31$ inclusive}{
					Add circuit from \cref{fig:circ:gimli_T_k} correspondingly $j,k$ and $L$\linebreak with $x\sim s_{0,j}, y\sim s_{1,j}$ and $z\sim s_{2,j}$.\;
				}
				Relabel qubit registers/update $L$ correspondingly \linebreak $s_{0,j}s_{1,j}s_{2,j}\leftarrow s_{2,j}s_{1,j}s_{0,j}$.\;
			}
			\If(\COMMENT*[f]{Small-Swap}){$r\equiv 0 \mod 4$}{
				Relabel qubit registers/update $L$ correspondingly \linebreak $s_{0,0}s_{0,1}s_{0,2}s_{0,3}\leftarrow s_{0,1}s_{0,0}s_{0,3}s_{0,2}$.\;
			}
			\If(\COMMENT*[f]{Big-Swap}){$r\equiv 2 \mod 4$}{
				Relabel qubit registers/update $L$ correspondingly \linebreak $s_{0,0}s_{0,1}s_{0,2}s_{0,3}\leftarrow s_{0,2}s_{0,3}s_{0,0}s_{0,1}$.\;
			}
			\If(\COMMENT*[f]{Add constant}){$r\equiv 0 \mod 4$}{
				Add \textsc{NOT} gates related to $bin(c)$, $bin(r)$ and $L$. \;
			}
		}
		\KwRet{$Q_\textsc{Gimli}, L$.}	
	}
	\caption{Quantum-\textsc{Gimli} circuit builder\label{alg:circ:q_gimli}.}
\end{algorithm}

%
%